\documentclass{wiley-article}

 \usepackage[super,sort&compress]{natbib}
\usepackage{siunitx}
\usepackage{multirow}
\usepackage{multicol}
\usepackage{xcolor}
\usepackage{booktabs}
\usepackage{colortbl}
\usepackage{physics}
\usepackage{parskip}

\papertype{Research Article}

\title{A polarizable CASSCF/MM approach using the interface between OpenMMPol library and CFour}

\author[1]{Tommaso Nottoli}
\author[1]{Mattia Bondanza}
\author[1]{Filippo Lipparini}
\author[1]{Benedetta Mennucci}

\affil[1]{Dipartimento di Chimica e Chimica Industriale, Universit\`a di Pisa, Via G. Moruzzi 13, I-56124 Pisa, Italy}

\corraddress{Dipartimento di Chimica e Chimica Industriale, Universit\`a di Pisa, Via G. Moruzzi 13, I-56124 Pisa, Italy}
\corremail{tommaso.nottoli@dcci.unipi.it; benedetta.mennucci@unipi.it}

\fundinginfo{}

\runningauthor{Nottoli et al.}

\begin{document}

\begin{frontmatter}
\maketitle

\begin{abstract}
\sloppy
We present a polarizable embedding quantum mechanics/molecular mechanics (QM/MM) framework for ground- and excited-state Complete Active Space Self-Consistent Field (CASSCF) calculations on molecules within complex environments, such as biological systems. These environments are modeled using the AMOEBA polarizable force field. This approach is implemented by integrating the OpenMMPol library with the CFour quantum chemistry software suite. The implementation supports both single-point energy evaluations and geometry optimizations, facilitated by the availability of analytical gradients. We demonstrate the methodology by applying it to two distinct photoreceptors, exploring the impact of the protein environment on the structural and photophysical properties of their embedded chromophores.

\end{abstract}

\end{frontmatter}

\section{Introduction}

Multiscale approaches that couple quantum mechanical (QM) descriptions with classical models have become a widely used strategy for investigating properties and processes of molecular systems in increasingly complex environments. Among these hybrid QM-classical methods, a particularly effective approach is the use of atomistic models based on Molecular Mechanics (MM) force fields for the classical component.\cite{Thiel_ACIE_QMMMRev,vanderKamp:2013em,Warshel:2014fz,Rothlisberger_ChemRev,Sousa:2016dma,Morzan2018} The success of the resulting QM/MM approach is largely attributed to its versatility. Any embedded system can, in principle, be treated at the QM/MM level, given the availability of an appropriate MM force field for the classical region and a model that accurately describes QM-MM interactions.
The nature of these QM-MM interactions distinguishes the various formulations of the QM/MM approach. Currently, two primary approaches are widely used: the standard electrostatic embedding and the more sophisticated polarizable embedding. The key difference between these methods is that polarizable embedding accounts for mutual polarization between the QM and MM regions, in addition to the electrostatic interactions. This is expected to play a significant role especially in processes where changes in the QM charge density are important such as in reactivity and light-induced phenomena.\cite{Bondanza:2020he,LocoAccount}

Polarizable embedding QM/MM models are available in various formulations, where the polarization of the MM component is represented using methods such as fluctuating charges,\cite{Rappe_JCP_QEq,Steven_JCP_MMPolFQ,lipparini_FQ,Poier_JCTC_BCFQ} Drude oscillators,\cite{Boulanger:2012gx,Riahi_JCC_QMDrudeMD,Lu_JCTC_QMDrude,ReviewDrude} or induced point dipoles\cite{Warshel1976,Applequist_JACS_PPD,Gao:1997de,vanduijnen98,curutchet_2009_electronicenergytransfer,Olsen_AQC_PE,Loco_JCTC_QMAMOEBA,Skylaris_JCTC_2,Wu:2017jg,Bondanza:2020he,nottoliwires}. The growing adoption of these models has been facilitated by the integration of these formulations into electronic structure codes or through interfaces between these codes and external libraries.\cite{mimic,LIBEFP,CPPE,bondanza2024ommp} Notably, the latter strategy has proven highly effective, as it enables a straightforward coupling of the polarizable embedding model with different QM codes and levels of theory.

Recently, our group has presented the OpenMMPol, an opensource library for coupling QM codes with induced point dipole (IPD)-based polarizable MM models.\cite{bondanza2024ommp}
OpenMMPol offers a modular implementation of all the functionalities to perform polarizable embedding calculations using either the standard IPD force-field (often referred as MMPol)\cite{Wang2011} or the more involved AMOEBA\cite{ponder2010current} force-field. Thanks to a simple but flexible multi-languange interface, OpenMMPol can easily work with most of existing QM softwares; the library also implements all the purely MM terms of the forcefield, including a simple and straightforward way to handle link-atoms and all the energy (and geometrical gradients) for the purely MM terms required by AMOEBA and AMBER forcefields. Thanks to those features, OpenMMPol is an appealing plug-and-play solution that allows, after a minimal interfacing effort, to perform complex QM/MM calculations on a broad variety of real-life systems without any external software required.

Here, the OpenMMPol library is interfaced with \textsc{CFour},\cite{cfour,matthews2020cfour} a suite of QM programs that specializes in highly accurate post-Hartree Fock calculations, including a large manifold of methods from the Coupled Cluster hierarchy, but also the Complete Active Space - Self-consistent field (CASSCF) method.\cite{roos1980,Werner1987,Shepard1987} The latter can be used for both ground- and excited-state calculations\cite{Lipparini2016,Lipparini2017} and, thanks to a recent implementation\cite{Nottoli2021,Nottoli22} based on the Cholesky Decomposition of the two-electron integrals\cite{Beebe1977,Roeggen2008,Aquilante2011,Koch2003}, it can be applied to large molecular systems.

CASSCF methods have been previously integrated with polarizable force fields.\cite{li2015polarizable,Liu:2019fna,Song23,Song24} In this work, the \textsc{CFour}-OpenMMPol interface enables an efficient implementation of the CASSCF/AMOEBA approach, allowing for the calculation of ground and excited state energies in both State-Average and State-Specific formulations. Additionally, this implementation supports geometry optimizations using analytical gradients.

\begin{figure}
    \centering
    \includegraphics[width=1.0\linewidth]{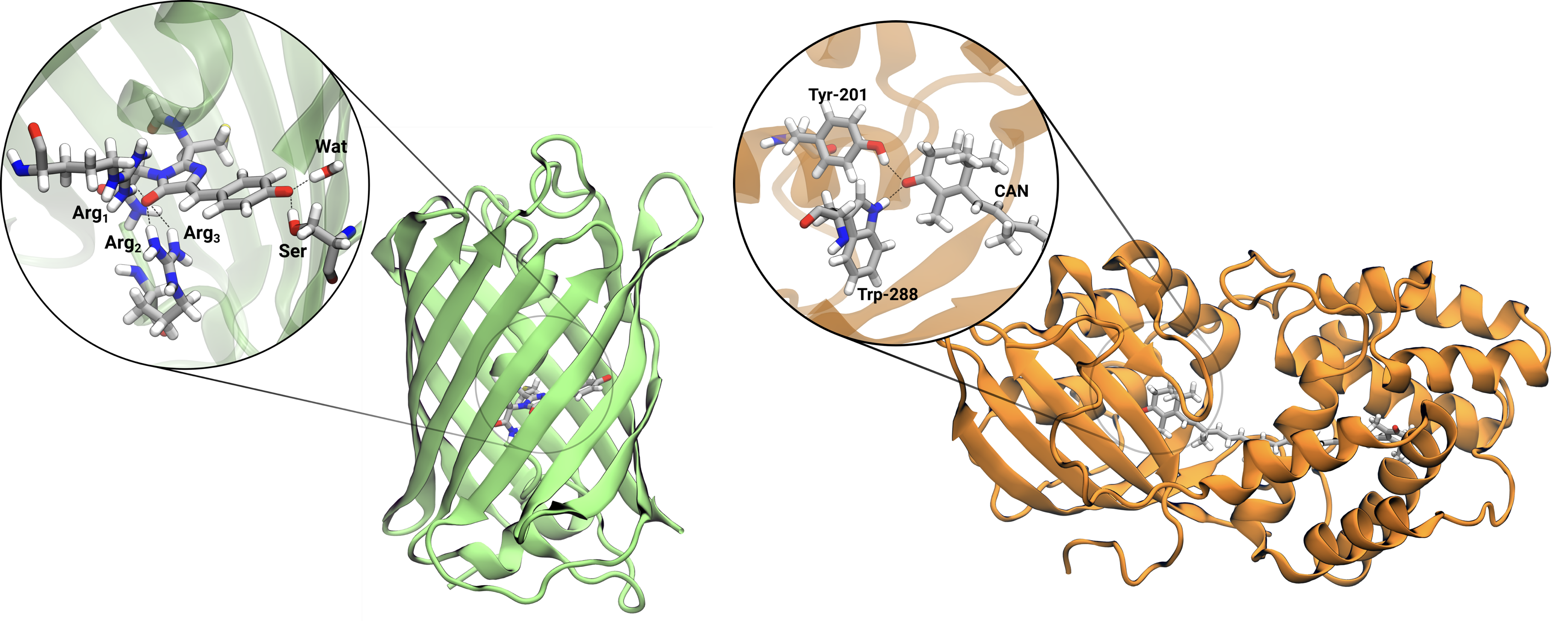}
    \caption{Representation of  Dronpa (left) and OCP (right) structures. The insets show the hydrogen bond networks of each chromophore in its binding pocket:  Serin-142 (Ser), Arginine-91 (Arg$_1$), two hydrogen atoms of Arginine-66 (Arg$_2$ and Arg$_3$), and a water molecule (Wat) for Dronpa and Tyrosine-201 (Tyr-201) and Tryptophan-288 (Trp-288) for OCP.}
    \label{fig:dronpa1ocp}
\end{figure}

The developed software was used to investigate two photoresponsive proteins (see Figure~\ref{fig:dronpa1ocp}). The first system is the Dronpa variant of green fluorescent protein (GFP), in which the chromophore spontaneously forms from the condensation of a tyrosate residue with adjacent glycine and cysteine. The second system is the orange carotenoid protein (OCP), where the chromophore is canthaxanthin (CAN). In both systems the chromophores are involved in significant hydrogen-bonding interactions with nearby residues. However, in Dronpa, the chromophore is covalently bonded to the protein matrix and carries a charge, whereas the CAN chromophore in OCP is noncovalently bound and is neutral.
These characteristics make the two systems valuable test cases for the CASSCF/AMOEBA approach. They allow for the investigation of both  specific hydrogen-bonding interactions and nonspecific electrostatic and polarization effects exerted by the protein on the chromophore geometry and response to light. To reach a deeper understanding of these effects, we also compare the AMOEBA force-field with an electrostatic embedding model.

\section{Method \& Implementation}

A convenient way of deriving a generic polarizable embedding model coupled to a QM method consists in building a Lagrangian.\cite{nottoli2020general} 
Adopting the AMOEBA force field, where a fixed multipolar distribution (up to quadrupoles) is used in combination with induced dipole moments, the Lagrangian reads:\cite{nottoliwires}
\begin{equation}
    \label{eq:LAmo}
    \mathcal{L}(\kappa,c,\mu^d,\mu^p) =  \mathcal{E}^{\rm CAS}(\kappa,c) + \mathcal{E}^{\rm self}(M) + \mathcal{E}^{\rm ele}(\kappa,c,M) 
    + \mathcal{E}^{\rm pol}(\kappa,c,M) +\frac{1}{2}\langle \mu^p,\mathbf{T}\mu^d - \mathbf{E}(\kappa,c) - \mathbf{E}^d(M)\rangle
\end{equation}
where the first term, $\mathcal{E}^{\rm CAS}(\kappa, c)$ is the CASSCF energy and the other four terms represent the QM-AMOEBA specific terms. We recall that AMOEBA, due to different screening rules used in the polarization step and in energy calculations, introduces two independent sets of induced dipoles that are called `polarization' ($\mu^p$) and `direct' ($\mu^d$).\cite{ponder2010current} 

$\mathcal{E}^{\rm CAS}(\kappa, c)$ only depends on the QM parameters, \emph{i. e.} orbital rotation ($\kappa$) and configurational coefficients ($c$) and it can be written in terms of one- and two-body reduced density matrices and integrals
\begin{equation}
    \label{eq:ene_qm}
    \mathcal{E}^{\rm CAS} = \sum_i\left(h_{ii} + F^I_{ii}\right) + \sum_{uv}\gamma_{uv}F^I_{uv} + \frac{1}{2}\sum_{uvxy}\Gamma_{uvxy}(uv|xy) + \mathcal{E}^{\rm nuc}
\end{equation}
where $h_{ii}$ are conventional one-electron integrals, $(uv|xy)$ are two-electron integrals written using Mulliken convention, $\gamma_{uv}$ and $\Gamma_{uvxy}$ are the one- and two-body reduced density matrices, respectively, $\mathcal{E}^{\rm nuc}$ is the nuclear repulsion energy, and $\mathbf{F}^I$ is the inactive Fock matrix.

In eq.~\eqref{eq:LAmo}
\begin{equation}
    \mathcal{E}^{\rm self}(M) = \frac{1}{2}\sum_{k\ne l=1}^{N_{\rm MM}} s_{kl}^m \sum_{L,L'=0}^2 M_k^L T_{kl}^{LL'}M_l^{L'}
\end{equation}
is the self-interaction energy of the AMOEBA multipolar distribution ($M^L$), where $L=0, 1, 2$ for charges, dipoles, and quadrupoles, respectively. $s^m_{kl}$ is a screening factor used to remove unwanted interactions and
\begin{equation}
    T_{kl}^{LL'} = \frac{\partial^L}{\partial r_k^L}\frac{\partial^{L'}}{\partial r_l^{L'}} \frac{1}{|r_k - r_l|}
\end{equation}
is the generalized Coulomb kernel.\cite{thole1981molecular}

The second and third terms of eq.~\eqref{eq:LAmo} are the electrostatic and polarization interactions between the QM and the MM components.
The electrostatic interaction energy, $\mathcal{E}^{\rm ele}$, between the MM multipoles and the QM charge density can be written as 
\begin{equation} 
    \label{eq:EQM-M}
    \mathcal{E}^{\rm ele}(\kappa,c,M) = \sum_{k}\sum_L M_k^L \left ( \sum_{\alpha} T^{L0}_{k\alpha} Z_{\alpha} - \sum_{pq}\gamma_{pq} \langle \phi_p | \hat{t}^L_{k} |\phi_q\rangle \right )
\end{equation}
where $\phi_p$ is a molecular orbital, and 
\begin{equation}
    \label{eq:top}
    \hat{t}^L_k = \frac{\partial^L}{\partial r_k^L} \frac{1}{|r_k - r|}
\end{equation}

The polarization energy can be written as the scalar product between the induced dipoles ($\mu^d$) and the sum of of the QM electric field ($\mathbf{E}(\kappa,c)$) and the \emph{polarization} field due to the multipolar distribution
\begin{equation}
    \mathcal{E}^{\rm pol}(\kappa,c,M) =- \frac{1}{2}\langle \mu^d,\mathbf{E}(\kappa,c) + \mathbf{E}^p(M)\rangle 
\end{equation}

\begin{equation}
    \label{eq:Ep}
    E_k^p(M) = - \left (\sum_{l} s^p_{kl} \sum_L \mathcal{T}^{1L}_{kl} M_l \right )
\end{equation}
with
\begin{equation}
\mathcal{T}_{kl}^{LL'} = \frac{\partial^L}{\partial r_k^L}\frac{\partial^{L'}}{\partial r_l^{L'}} \frac{\lambda(|r_k-r_l|)}{|r_k - r_l|}
\end{equation}
as the damped Coulomb kernel, which is used to avoid the so-called polarization catastrophe.\cite{thole1981molecular,wang2011development,sala2010polarizable} 

The coupled CASSCF/AMOEBA equations can be obtained by imposing the stationarity conditions on the Lagrangian in eq.~\eqref{eq:LAmo}. 

By differentiating with with respect to $\kappa_{pq}$, we get the CASSCF orbital gradient 
\begin{equation}
    \label{eq:cas_egrad}
    g_{pq} = 2\left(F_{pq} - F_{qp}\right)
\end{equation}
where $F$ is the generalized Fock matrix which contains a modified inactive Fock matrix
\begin{equation}\label{eq:fock_env}
    \Bar{F}^I_{pq} = F^I_{pq} - \sum_k^{N_{\rm MM}}\sum_LM^L_k\bra{\phi_p}\hat{t}^L_k\ket{\phi_q} - \sum_k^{N_{\rm MM}}\frac{\mu^d_k+\mu^p_k}{2}\bra{\phi_p}\hat{t}^1_k\ket{\phi_q},
\end{equation}
which includes the effects of the environment as additional one-electron contributions. 
Here we exploited a nested, first-order, algorithm to get the optimal CASSCF parameters. Therefore, the full CI problem is solved at each orbital optimization step up to a numerical threshold, which depends on the norm of the orbital gradient. In particular, we used the Super-CI algorithm as the orbital optimization solver.\cite{roos1980,siegbahn1981,kollmar2019,angeli2002}

Differentiation eq.~\eqref{eq:LAmo} with respect to $\mu^d$ and $\mu^p$ leads, respectively, to the following polarization equations that need to be solved at each CASSCF step
\begin{align}
    &\mathbf{T}\bm{\mu}^d = \mathbf{E}(\kappa,c) + \mathbf{E}^d(M)\\
    &\mathbf{T}\bm{\mu}^p = \mathbf{E}(\kappa,c) + \mathbf{E}^p(M) 
\end{align}
where the $kl-$th block of matrix $\mathbf{T}$ is defined as follows
\begin{equation}
    T_{kl} = \alpha_k^{-1}\delta_{kl} + (1-\delta_{kl}) \mathcal{T}_{kl}^{11}
\end{equation}
and $\alpha_k$ is the polarizability of the $k$-th MM atom. 
Here $\mathbf{E}^d(M)$ is the \emph{direct} field 
\begin{equation}
    \label{eq:Ed}
    E_k^d(M) = - \left (\sum_{l} s^d_{kl} \sum_L \mathcal{T}^{1L}_{kl} M_l \right )
\end{equation}
which differs from the polarization field reported in eq.~\eqref{eq:Ep} because the screening factors are different.

The OpenMMPol-\textsc{CFour} interface works as follows. The electrostatic contribution to the Fock matrix (second term of eq.~\eqref{eq:fock_env}) is computed at the beginning by contracting the appropriate integrals in the AO basis with the multipoles, which can be accessed by the OpenMMPol object. At each CASSCF optimization step, the one-body reduced density matrix is used to compute the electronic component of the electric field and passed as input to the OpenMMPol subroutine \texttt{set\_external\_field}, which in turn solves the polarization equations. Successively, the induced dipoles are used to compute the third term of eq.~\eqref{eq:fock_env}. 

Finally, the geometrical gradients can be obtained by differentiating eq.~\eqref{eq:LAmo} with respect to the atomic coordinates. Moreover, only explicit derivatives need to be computed for all the contributions. 
We consider here only the gradient with respect to QM coordinates ($r_\alpha$) and for a single CASSCF state. 
Differentiation of the first term of eq.~\eqref{eq:LAmo} leads to the standard CASSCF molecular gradient. The other contributions stem from the electrostatic interaction 
\begin{equation}
    \pdv{\mathcal{E}^{\rm ele}}{r_\alpha}(\kappa,c,M) = \sum_{k}\sum_L M_k^L \left ( T^{L1}_{k\alpha} Z_{\alpha} - \sum_{pq}\gamma_{pq} \pdv{\langle \phi_p | \hat{t}^L_{k} |\phi_q\rangle}{r_\alpha} \right ),
\end{equation}
and from the polarization energy
\begin{equation}
    \pdv{\mathcal{E}^{\rm pol}}{r_\alpha} = - \frac{1}{2}\sum_k \left( \mu^d_k + \mu^p_k\right) \pdv{E_k(\kappa,c)}{r_\alpha}.
\end{equation}

The OpenMMpol-CFour interface can also be used for computing excitation energies.
Our method is based on state-averaged CASSCF and allows for the description of several excited states. During the CASSCF procedure, the environment polarization is converged with respect to the state-averaged density of the QM system.\cite{Song_CASAMOEBA} 
In order to recover a state-specific polarization of the environment for each excitation, a correction is introduced following the approach proposed by Li \emph{et al.}\cite{li2015polarizable}.
At convergence of the state-average calculation, for the two states involved in the excitation, we compute the induced dipoles using the corresponding state density  ($\gamma^X$) and we recompute the electrostatic and polarization terms of the energy 
\begin{equation}
         \mathcal{E}_{\rm corr}^X = \mathcal{E}^{\rm ele-pol}(\gamma^X, \mu(\gamma^X)) - \mathcal{E}^{\rm ele-pol}(\gamma^X, \mu(\Bar{\gamma}))
\end{equation}
Finally, we correct the state-averaged excitation energy ($\Delta\mathcal{E}_{\rm SA}^{IF}$) to get the state-specific corrected one as it follows:
\begin{align}
    \Delta\mathcal{E}_{\rm SS}^{IF} &= \Delta\mathcal{E}_{\rm SA}^{IF} + \mathcal{E}_{\rm corr}^{IF}\\
    \Delta\mathcal{E}_{\rm corr}^{IF} &= \mathcal{E}_{\rm corr}^F - \mathcal{E}_{\rm corr}^I
\end{align}
where $I$ and $F$ indicate the initial and the final state involved in the excitation, respectively.


\section{Two applications}

In this section, we report the results of the CASSCF/AMOEBA description obtained via the \textsc{CFour}-OpenMMPol interface to study the impact of the protein environment on the geometry and excitation energies of the two chromophores within Dronpa and OCP. To reach a deeper understanding of these effects, we compare the AMOEBA force-field with the commonly utilized electrostatic embedding model (specifically, AMBER99SB force-field), where the environment is described using the Amber force field. Furthermore, we assess the performance of CASSCF/AMOEBA relative to the (TD)DFT level of theory.

The starting structures for the optimization procedure, are taken from previous works on OCP\cite{Bondanza2020} and Dronpa\cite{Nifos2019} by our group and have been prepared in a similar way: the crystal structure (2Z1O for Dronpa\cite{dronpapdb} and 4XB5 for OCP\cite{Leverenz:2015jaa}) was inserted in a solvent box containing water molecules and enough Na$^+$ and Cl$^-$ ions to obtain a neutral system and a salt concentration of 0.1 M. The details of the equilibration procedures are explained in the reference papers. In our optimizations only solvent molecules within 5~\AA ~distance from the protein are included. 
 
All CASSCF calculations were performed using unrestricted natural orbitals (UNO) as the initial guess. In addition to being well suited for the CASSCF optimization algorithm, UNOs allow for automatic determination of the active space based on natural occupation numbers.\cite{pulay1988uhf,toth2020comparison} The optimal strategy is to include all orbitals with occupation numbers (ON) between 0.01 and 1.99. However, if this results in an excessively large active space, it is reduced to ensure that the calculation remains feasible. In both systems, the QM region includes only the chromophore.
All CASSCF geometry optimizations were performed with the MM atoms kept frozen and the 6-31G(d) basis set,\cite{hehre1972_631gd} whereas the 6-31+G(d) was employed for excited-state calculations. Various active spaces were tested (see SI), but the results presented here are based on a (12,12) space for Dronpa and a (8,12) space for OCP. Specifically, for Dronpa we included all orbitals as provided by the UNO criterion plus additional $\pi$ ones, while for OCP we considered as active the orbitals with ON between 0.04 and 1.93. 
For the SA-CASSCF calculations, we considered two states for Dronpa, with the exception of the Amber calculation for which three states were averaged to improve the convergence. On the other hand, three states were always considered for OCP.
For the (TD)DFT calculations, the B3LYP exchange correlation functional\cite{B3LYP} in conjunction with the 6-31G(d) basis set was used for geometry optimization, while the CAM-B3LYP functional\cite{CAMB3LYP} was applied, using the 6-31+G(d) basis, to compute vertical excitations. 

\subsection{Dronpa}

\begin{figure}
    \centering
    \includegraphics[width=0.4\linewidth]{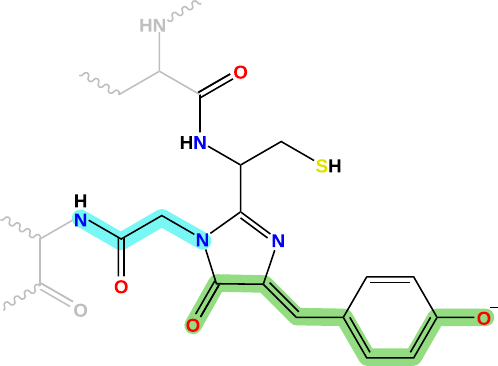}
    \caption{Chemical representation of the Dronpa chromophore. In gray we show the residues to which it is covalently bonded, which are treated at the MM level either using the Amber or AMOEBA force field. The atoms considered to compute the BLA are highlighted in green, whereas those that were kept frozen during the optimizations performed in vacuum are highlighted in cyan.}
    \label{fig:dronpa2}
\end{figure}

The chromophore of Dronpa can be described in terms of two resonance structures, depending on the formal position of the negative charge. In the benzenoid form, depicted in Figure~\ref{fig:dronpa2}, the charge is primarily localized on the oxygen atom attached to the benzene ring, while in the quinoid form, the charge resides on the oxygen atom of the imidazole ring.
To investigate the effect of the protein in tuning charge distribution thus affecting the weight of the two limiting resonance structures, we use the bond length alternation (BLA) parameter. BLA is defined as the difference between the average lengths of the single and double bonds highlighted in green in Figure~\ref{fig:dronpa2}. A positive BLA value indicates a predominance of the benzenoid form, while a negative BLA signifies the quinoid form.

 In Table~\ref{tab:bla_dihed_dronpa}, we report the values of the BLA computed for the isolated chromophore (vacuum) and in the protein with the two selected force fields. For the isolated chromophore, the optimization was performed with the dihedral angle that connects the nitrogen of the imidazole ring to the nitrogen of the condensed glycine (see atoms highlighted in cyan in Figure~\ref{fig:dronpa2}) frozen to prevent the formation of a hydrogen bond between the oxygen of the imidazole ring and the amide hydrogen. The dihedral angle was fixed to the value found in the protein cavity. 
   
\begin{table}[h]
\caption{DFT and CASSCF BLA values (in \AA) and  dihedral angles (in degrees) for Dronpa's chromophore computed in  vacuum and in protein with the two force fields. }
\label{tab:bla_dihed_dronpa}
\begin{threeparttable}
\begin{tabular}{lccc}
\headrow
\thead{Method} & \thead{Vacuum} & \thead{Amber} & \thead{AMOEBA} \\
\hiderowcolors
\multicolumn{4}{c}{BLA} \\
B3LYP                & 0.014     & 0.004     & 0.016     \\
CAS(12,12)           & 0.041     & 0.010     & -0.020    \\
\hline  
\multicolumn{4}{c}{Dihedral Angle} \\
B3LYP                & 179      & 174      & 175      \\
CAS(12,12)           & 180      & 174      & 174      \\
\hline  
\end{tabular}
\end{threeparttable}
\end{table}

CASSCF and DFT give a qualitatively similar description for both the isolated chromophore and the chromophore embedded in an electrostatic embedding (Amber). In contrast, the two QM methods show qualitative differences when using the polarizable AMOEBA force field. Specifically, DFT/AMOEBA yields a value similar to the vacuum case, while CASSCF/AMOEBA gives a negative BLA indicative a quinoid-like character.

Additional insights can be gained by examining the dihedral angle around the single bond adjacent to the benzene ring, as shown in Table~\ref{tab:bla_dihed_dronpa}.
The dihedral angle is smaller in protein than in vacuum, where the molecule adopts a planar conformation, but it shows little sensitivity to the environmental model used. For both force fields, the dihedral angle remains nearly identical (around 174°). This tiny distortion of the chromophore appears to be primarily driven by the steric effects of the protein pocket. The discrepancy between the embedding models in describing the BLA, cannot therefore be attributed to the distortion of the dihedral angle, but is driven by local interactions, such as hydrogen bonds, present within the protein.
The two oxygen atoms at the extremities of the conjugated chain considered in the BLA calculation in fact engage in multiple hydrogen bonds with nearby residues. The hydrogen bond network within the protein cavity is illustrated in the inset of Figure~\ref{fig:dronpa1ocp}.  

Table~\ref{tab:Hbonds_dronpa} presents the lengths of the key hydrogen-bonds between the oxygen atom of the benzene ring and Serine-142 (Ser), as well as a water molecule (Wat), and between the oxygen atom of the imidazole ring and Arginine-91 (Arg$_1$), alongside two hydrogen atoms of Arginine-66 (Arg$_2$ and Arg$_3$). Moving from the electrostatic to the polarizable embedding results a noticeable increase in the bond lengths is observed for both DFT and CASSCF methods. Specifically, CASSCF exacerbates these differences, with the most significant change being a 0.165~{\AA} increase in the hydrogen bond length with Arg$_3$ when switching from Amber to AMOEBA. This different description of H-bonds may explain the reversed BLA signs between Amber and AMOEBA computed at the CASSCF level.

\begin{table}[h]
\caption{DFT and CASSCF values of the H-bond lengths between Dronpa's chromophore and the residues/water computed with the two force fields. All values are in \AA. 
}
\label{tab:Hbonds_dronpa}
\begin{threeparttable}
\resizebox{\textwidth}{!}{
\begin{tabular}{lcccccccccc}
\hiderowcolors
\rowcolor[gray]{0.80}
    & \multicolumn{2}{c}{\textbf{Ser}} & \multicolumn{2}{c}{\textbf{Wat}} & \multicolumn{2}{c}{\textbf{Arg$_1$}} & \multicolumn{2}{c}{\textbf{Arg$_2$}} & \multicolumn{2}{c}{\textbf{Arg$_3$}}\\  
\textbf{Method} & Amber & AMOEBA & Amber & AMOEBA & Amber & AMOEBA & Amber & AMOEBA & Amber & AMOEBA\\  
\hiderowcolors
B3LYP      &    1.821 &    1.857 &    1.814 &    1.819 &    1.932 &    1.934 &    2.433 &    2.524 &    1.826 &    1.907 \\
CAS(12,12) &    1.751 &    1.787 &    1.742 &    1.828 &    1.898 &    1.970 &    2.455 &    2.508 &    1.787 &    1.952 \\
\hline
\end{tabular}
}
\end{threeparttable}
\end{table}

The effect of the embedding model has also been assessed by computing excitation energies. In order to disentangle the impact of the geometry, for AMOEBA calculations of excitation energies, we performed two calculations using Amber (AMOEBA@Amber) and AMOEBA (AMOEBA@AMOEBA) optimized geometries. 

Table~\ref{tab:eexc_dronpa} presents the first excitation energy calculated at the SA-CAS(12,12)/6-31+G(d) and TDCAM-B3LYP/6-31+G(d) level on CASSCF geometries for the two embedding models and in vacuum. For CAS/AMOEBA calculations we report also the state-specific corrected energies ($\Delta \mathcal{E}_{\rm SS}$). To have a more correct comparison with $\Delta \mathcal{E}_{\rm SS}$ for TDDFT/AMOEBA calculations we report the values obtained within a State-Specific approximation generally indicated as ``corrected linear response'' cLR.\cite{Loco_JCTC_QMAMOEBA}

\begin{table}[h]
    \caption{Excitation energies (in eV) of the first excited state of Dronpa's chromophore computed using SA-CAS(12,12) and TDCAM-B3LYP level on top of CAS geometries. For CAS/AMOEBA calculations we report also the state-specific corrected energies ($\Delta \mathcal{E}_{\rm SS}$). For TDDFT calculations we also report the excitation energies calculated using DFT optimized geometries (TDDFT/DFT). The notation AMOEBA@Amber and AMOEBA@AMOEBA means that the excitation energy has been computed with  QM/AMOEBA at the geometry optimized with QM/Amber and QM/AMOEBA model, respectively.}
    \label{tab:eexc_dronpa}
    \begin{threeparttable}
    \begin{tabular}{c|cc|cc}
    \headrow
Method & \multicolumn{2}{|c|}{SA-CAS/CAS} & TDDFT/CAS & TDDFT/DFT\\
\hline
 & $\Delta \mathcal{E}_{\rm SA}$ & $\Delta \mathcal{E}_{\rm SS}$ & $\Delta \mathcal{E}$ & $\Delta \mathcal{E}$\\
\hline
\hiderowcolors
        Vacuum    & 3.41  &      & 3.15 & 3.04 \\
        Amber      & 3.48  &      & 3.13 & 3.22 \\
        AMOEBA@Amber     & 3.63  & 3.65 & 3.22 &  3.30 \\
        AMOEBA@AMOEBA    & 3.50  & 3.58 & 3.24 & 3.16 \\
        \hline
    \end{tabular}
    \end{threeparttable}
\end{table}

Starting from the CAS values, we observe that by keeping fixed the model used to compute the excitation energy, the effect of a different geometry is not negligible: $\Delta\mathcal{E}_{\rm SA}$ of AMOEBA@AMOEBA is red-shifted by 0.13 eV with respect to the AMOEBA@Amber case. However, this difference is reduced to 0.07 eV when the state-specific correction is added to the AMOEBA calculations.  
Instead, by keeping constant the Amber geometry and changing the model for the energy calculations, we note that AMOEBA and Amber excitation energies differ by 0.17 eV. 
Comparing Vacuum, Amber, and AMOEBA@AMOEBA, we observe a systematic increase in the excitation energy moving from the isolated molecule to the protein with an overall blue shift of 0.17 eV for AMOEBA of which 0.1 eV is its difference with respect to Amber. If we do not include the state-specific correction, the latter difference almost disappears making the two embedding models very similar.  
All these data clearly show that including a SS correction for the polarizable model is really important. In this case, it can also be justified by a somewhat large difference between the ground- and excited-state dipole, 17 and 11 Debye, respectively.

Moving to the energies computed at the TDDFT level on the same CAS geometries and on the DFT geometries, both qualitative and quantitative variations are observed.
The excitation energy is red-shifted relative to the CAS value for the isolated molecule using the same geometry, with an additional red shift observed when using DFT geometry. Considering environmental effects, further differences emerge. 
The behavior we obtain at TDDFT when moving from vacuum to AMOEBA is qualitatively similar with respect to what found at CAS level if we consider consistent calculations (e.g. AMOEBA@AMOEBA): for both CAS and DFT geometries we observe a blue-shift of about 0.1 eV. In contrast, the Amber-AMOEBA shift strongly depends on the selected geometry and shows a blue-shift for CAS geometries and a red-shift for DFT geometries.
All these data show a larger sensitivity of the TDDFT description to geometries.

\subsection{OCP}

The analysis presented for Dronpa is here repeated for OCP.

\begin{figure}[h]
    \centering
    \includegraphics[width=0.5\linewidth]{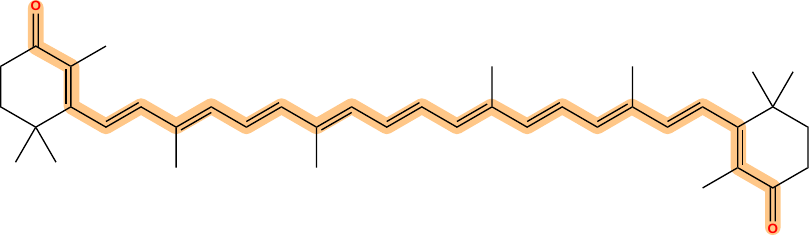}
    \caption{Chemical representation of the CAN chromophore. The atoms considered to compute the BLA are highlighted.}
    \label{fig:can}
\end{figure}

As before we start the analysis by presenting the calculated values of the BLA parameter for CAN. In this case, we included all atoms involved in the conjugated chain of CAN, as depicted in Figure~\ref{fig:can}. 

The results obtained at CAS and DFT level of theories and with the different embedding models are reported in Table~\ref{tab:bla_ocp}.

At the DFT level, both force fields exhibit nearly identical BLA. However, CASSCF results show a modest increase in BLA when using AMOEBA. Overall, CASSCF tends to overestimate the BLA compared to DFT -- which have been shown to be consistent with higher levels of theory;\cite{bondanza2021excited} this effect can be attributed to the absence of dynamic correlation in CASSCF -- a critical factor in the geometry optimization of conjugated molecules.\cite{peach2007structural,bondanza2021excited}

\begin{table}[h]
\caption{CAS and DFT BLA values (in \AA) for CAN computed in vacuum and in protein with the two force fields.}
\label{tab:bla_ocp}
\begin{threeparttable}
\begin{tabular}{lccc}
\headrow
\thead{Method} & \thead{Vacuum} & \thead{Amber} & \thead{AMOEBA} \\
\hiderowcolors
B3LYP                & 0.110     & 0.101     & 0.102     \\
CAS(8,12)            & 0.151      & 0.147     & 0.149     \\
\hline  
\end{tabular}
\end{threeparttable}
\end{table}

As reported in the Introduction, CAN is engaged in two hydrogen bonds with two adjacent residues, namely a tryptophan (Trp-288) and a tyrosine (Tyr-201). 
In Table~\ref{tab:Hbonds_ocp} we report the corresponding bond lengths as obtained through CAS and DFT optimizations using either Amber of AMOEBA. As already found for Dronpa, we observe a general increase on the distances when moving from a non-polarizable to a polarizable force field, both for DFT and CASSCF data. The hydrogen bonds computed at the DFT and CASSCF level formed with Trp-288 reasonably agree. On the other hand, we note a somewhat higher difference of up to 0.05~{\AA} for Tyr-201.

\begin{table}[h]
\caption{DFT and CASSCF values of the H-bond lengths between OCP’s chromophore and the residues computed with the two force fields. All values are in {\AA}. 
}
\label{tab:Hbonds_ocp}
\begin{threeparttable}
\begin{tabular}{lcccc}
\hiderowcolors
\rowcolor[gray]{0.80}
    & \multicolumn{2}{c}{\textbf{Trp-288}} & \multicolumn{2}{c}{\textbf{Tyr-201}} \\  
\textbf{Method} & Amber & AMOEBA & Amber & AMOEBA \\  
\hiderowcolors
B3LYP      &    1.853 &   1.916 &   1.815 &   1.888 \\
CAS(8,12)  &    1.879 &    1.925 &    1.869 &    1.924 \\
\hline
\end{tabular}
\end{threeparttable}
\end{table}

Moreover, the directional nature of the hydrogen bond network can be assessed by looking at the angle formed between the oxygen atom of CAN and the hydrogen atoms of Trp-288 and Tyr-201. Here, the two QM levels of theory provide almost the same result, 90$^\circ$ with Amber and 88$^\circ$ for AMOEBA.

In Table~\ref{tab:eexc_ocp} we compare the excitation energy to the lowest bright state of CAN in OCP calculated using CASSCF and TDDFT, the latter performed either on top CASSCF optimized geometries or DFT optimized ones. 

\begin{table}[h]
    \caption{Excitation energies (in eV) of the first excited state of OCP's chromophore computed using SA-CAS(8,12)/6-31+G(d) and TDCAM-B3LYP/6-31+G(d) level on top of CASSCF(8,12)/6-31G(d) geometries. For CAS/AMOEBA calculations we report also the state-specific corrected energies ($\Delta \mathcal{E}_{\rm SS}$). For TDDFT calculations we also report the excitation energies calculated using B3LYP/6-31G(d) optimized energies (TDDFT/DFT). The notation AMOEBA@Amber and AMOEBA@AMOEBA means that the excitation energy has been computed with the QM/AMOEBA at the geometry optimized with the QM/Amber and QM/AMOEBA model, respectively.}
    \label{tab:eexc_ocp}
    \begin{threeparttable}
    \begin{tabular}{c|cc|cc}
    \headrow
Method & \multicolumn{2}{|c|}{SA-CAS/CAS} & TDDFT/CAS & TDDFT/DFT\\
\hline
 & $\Delta \mathcal{E}_{\rm SA}$ & $\Delta \mathcal{E}_{\rm SS}$ & $\Delta \mathcal{E}$ & $\Delta \mathcal{E}$\\
\hline
\hiderowcolors
        Vacuum    &   4.52    &      &   3.03   &   2.26  \\
        Amber      &   4.24    &      &   2.73   &   2.30  \\
        AMOEBA@Amber     &   4.30    &  4.35    &   2.89   &   2.31  \\
        AMOEBA@AMOEBA    &   4.50    &  4.59    &   2.93   &   2.36  \\
        \hline
    \end{tabular}
    \end{threeparttable}
\end{table}

The CASSCF-computed excitation energies are consistently blue-shifted relative to the TDDFT results, which is in line with expectations.\cite{helmich2019benchmarks} As observed for Dronpa, the impact of using different geometries for CASSCF excitation energies and the Amber force field for the environment is negligible. In contrast, AMOEBA@Amber and AMOEBA@AMOEBA differ by 0.24 eV, indicating greater sensitivity to molecular geometry with the AMOEBA force field.
Comparing consistent calculations (Vacuum, Amber and AMOEBA@AMOEBA), we find that introducing an electrostatic embedding causes a significant red shift (around 0.3 eV) from vacuum, but this shift is fully counteracted by AMOEBA.
This behavior is confirmed by the TDDFT/CAS results: a red shift of 0.3 eV is observed with Amber, which is largely reduced by AMOEBA. In contrast, the TDDFT/DFT results show a blue shift for both Amber and AMOEBA. At TDDFT level, the effect of the geometry on the AMOEBA excitation energies is much smaller than what found at CAS level.

\section{Conclusions}
We have presented an implementation of a polarizable embedding QM/MM that can be used to perform ground and excited state CASSCF calculations of molecules embedded in a polarizable environment described with the AMOEBA force field. The implementation has been achieved by coupling the recently developed OpenMMPol library\cite{bondanza2024ommp} and the CFour suite of quantum-mechanical programs. 
OpenMMPol offers a versatile and user-friendly realization of all the machinery required to run embedded calculations, including an implementation of all the purely MM contributions to the energy (\emph{e.g.}, bonded, van der Waals) and to the forces, plus electrostatic drivers that can be used to compute the induced point dipoles and the various QM/MM interaction terms. The coupling with the CASSCF program in CFour, which is well suited for calculations on medium and large systems thanks to an efficient implementation based on the Cholesky Decomposition of the two-electron integrals, opens the way to the description of complex molecular systems embedded in biological matrices, and is particularly useful to obtain a qualitative description of systems that exhibit a marked multireference character in their ground and excited states. 

This work opens the door to several future directions that can leverage the many capabilities of CFour to perform highly accurate calculations of molecular energies, structures, and properties. In particular, we plan to extend the present implementation to embedded Coupled-Cluster (CC) theory, including Equation of Motion CC (EOM-CC), and to complete the CASSCF-OpenMMPol interface to allow for linear response calculations of excitation energies, transition moments, and frequency dependent properties. We plan to include the interface described in this paper in the next public release of CFour.

\section*{Acknowledgments}
The authors acknowledge funding from ICSC-Centro Nazionale di Ricerca in High-Performance Computing, Big Data, and Quantum Computing, funded by the European UnionNextGenerationEU-PNRR, Missione 4 Componente 2 Investimento 1.4.

\section*{Conflict of Interest}
There are no conflicts of interest.

\section*{Supporting Information}
Additional data, using different CAS active spaces, on the geometrical parameters for the two systems are provided in the Supporting Information.

\bibliography{biblio}

\end{document}